\documentclass[asp,prl,amsmath,amssymb,reprint,superscriptaddress,preprintnumbers,showpacs,intlimits,shortbibliography]{revtex4-2}
\pdfoutput=1
\bibpunct{[}{]}{,}{n}{}{}

\usepackage[utf8]{inputenc}
\usepackage{bm,latexsym,mathrsfs,enumerate}
\usepackage[dvipsnames]{xcolor}
\usepackage{mathtools}
\usepackage{makecell}
\usepackage{multirow}
\usepackage[breaklinks=true,unicode=true,urlcolor = blue,colorlinks = true,citecolor = blue,linkcolor = blue]{hyperref}
\usepackage{graphicx}
\usepackage{ragged2e}
\usepackage[export]{adjustbox}
\usepackage[final]{pdfpages}
\usepackage{tikz}
\makeatletter
\AtBeginDocument{\let\LS@rot\@undefined}
\makeatother
\begin{document}
	
\title{Chirality-inverted Dzyaloshinskii--Moriya interaction}

\author{Khalil~Zakeri}
\email{khalil.zakeri@partner.kit.edu}
\affiliation{Heisenberg Spin-dynamics Group, Physikalisches Institut, Karlsruhe Institute of Technology, Wolfgang-Gaede-Str. 1, D-76131 Karlsruhe, Germany}
\author{Alberto Marmodoro}
\affiliation{Institute of Physics, Academy of Science of the Czech Republic, Cukrovarnická 10, Praha 6, CZ-16253, Czech Republic}
\author{Albrecht~von~Faber}
\affiliation{Heisenberg Spin-dynamics Group, Physikalisches Institut, Karlsruhe Institute of Technology, Wolfgang-Gaede-Str. 1, D-76131 Karlsruhe, Germany}
\author{Sergiy Mankovsky}
\affiliation{Department of Chemistry and Physical Chemistry, LMU Munich, Butenandtstrasse 11, D-81377 Munich, Germany}
\author{Hubert Ebert}
\affiliation{Department of Chemistry and Physical Chemistry, LMU Munich, Butenandtstrasse 11, D-81377 Munich, Germany}
\begin{abstract}
Dzyaloshinskii--Moriya interaction (DMI) is an antisymmetric exchange interaction, which is responsible for the formation of  topologically protected spin textures in  chiral magnets. Here by measuring the dispersion relation of the DM energy, we quantify the atomistic DMI in a model system, i.e., a Co double layer on Ir(001). We unambiguously demonstrate the presence of a  chirality-inverted DMI, i.e,  a sign change in the chirality index of DMI from negative to positive, when comparing the interaction between nearest neighbors to that between neighbors located at longer distances. The effect is in analogy to the change in the character of the Heisenberg exchange interaction from, e.g., ferromagnetic to antiferromagnetic. We show that the pattern of the atomistic DMI in epitaxial magnetic structures can be very complex and provide critical insights into the nature of DMI.
We anticipate that the observed effect is general and occurs in many magnetic nanostructures  grown on heavy-element metallic substrates.
\end{abstract}

\date{\today}
\maketitle

Exchange interaction is an inherently quantum mechanical effect, which describes the fundamental interaction between indistinguishable particles. In magnetic solids, the symmetric Heisenberg exchange interaction (HEI), is essential to understand the magnetic order \cite{Heisenberg1928}.
In a spin Hamiltonian representation, having the form of $\mathcal{H_{\mathrm{HEI}}}=-\sum_{i\neq j}J_{ij}\mathbf{S}_i \cdot\mathbf{S}_j$, the exchange coupling parameter $J_{ij}$, which describes the interaction between atomic spins $\mathbf{S}_i$ and $\mathbf{S}_j$, is symmetric with respect to permutation of the atoms  on sites $i$ and $j$. The ferro- or antiferromagnetic interaction manifests itself in the sign of $J_{ij}$.
Depending on the mechanism dominating the interatomic exchange interaction in a material, the coupling  can be either short-range, e.g., as in the case of the direct exchange or superexchange interactions, or long-range, as for instance in the case of the Ruderman-Kittel-Kasuya-Yosida  (RKKY) interaction \cite{Ruderman1954,Kasuya1956,Yosida1957},
having oscillatory behavior as a function of the interatomic distance. The latter one is intrinsic for the itinerant-electron systems and
may be significant in 3D systems, e.g., bcc Fe \cite{Kvashnin2016,Pajda2001,Halilov1998,Katsnelson2000,Szilva2013,Singer2011}, and in 2D systems, e.g., ultrathin films \cite{Zakeri2013a, Meng2014}. Moreover, $J_{ij}$ can change sign from positive (ferromagnetic) to negative (antiferromagnetic) as a function of the interatomic distance in an oscillatory manner \cite{Meng2014}.

In addition to HEI there exist another type of exchange interaction that is of antisymmetric nature. This interaction which has the form of $\mathcal{H_{\mathrm{DMI}}}=\sum_{i\neq j}\mathbf{D}_{ij}\cdot \mathbf{S}_i \times \mathbf{S}_j$ is known as Dzyaloshinskii--Moriya interaction (DMI) \cite{Dzyaloshinsky1958, Moriya1960}. Here the so-called DM vector $\mathbf{D}_{ij}$ is antisymmetric with respect to permutation of the sites $i$ and $j$. The direction of $\mathbf{D}_{ij}$  is governed by the symmetry of the lattice \cite{Moriya1960} and determines the twist of spins.
DMI is a chiral interaction and  has been shown to be a consequence of spin--orbit coupling (SOC) in spin systems with broken inversion symmetry \cite{Dzyaloshinsky1958, Moriya1960}.

In the case of ultrathin magnetic films, nanostructures as well as separated magnetic atoms deposited on heavy-element metallic substrates DMI can be significant due to the strong SOC of the substrate. However, since each of these systems belong to a different dimensionality class, different interactions act on individual atomic spins, and hence the microscopic nature of the interaction responsible for the DMI is different. It has been discussed that the DMI between the deposited individual magnetic atoms may change its sign as a function of the separation distance in an oscillatory manner \cite{Khajetoorians2016}. A similar behavior is predicted for magnetic chains deposited on stepped surfaces \cite{Schweflinghaus2016}. In the case of ultrathin magnetic films, multilayers and nanostructures the atomistic DMI vectors are linked to the symmetry of the underlying lattice and hence form an array of chiral vectors. In such a case the interesting feature would be that any change in the sign of the DMI vectors would lead to a chirality inversion within this array and a possible local change in the winding number $\mathcal{Q}$. Unfortunately, the experimental proof of any sign-change in the DMI vectors in such planar magnets has remained elusive, since a quantitative experimental determination of the atomistic DMI vectors is challenging. It is well-known that in 2D systems DMI leads to the formation of spin textures, e.g., magnetic helices \cite{Vedmedenko2007}, spirals \cite{Bode2007}, skyrmions \cite{Roessler2006,Heinze2011},  antiskyrmions \cite{Hoffmann2017}, merons \cite{Yu2018} and antimerons \cite{Hayami2021}, some of which are topologically protected. For an experimental design of a specific spin texture a detailed knowledge of the fundamental magnetic interactions in such 2D systems is of prime importance.  Hence, in the case of magnetic structures of itinerant
electron character grown on substrates with a strong SOC one may ask the following questions. (i) Is DMI in such systems similarly long ranged as HEI? (ii) Is the sign of $\mathbf{D}_{ij}$ only given by the symmetry of the lattice? (iii) Can the sign of $\mathbf{D}_{ij}$ change with interatomic distance, in a similar fashion as $J_{ij}$? If this is true, one would expect a chirality inversion of DMI when comparing the nearest neighbor interaction to the interaction between spins located at a longer distances.  This would provide guidelines for designing novel spin textures, e.g., atomic scale skyrmionium with  $\mathcal{Q}=0$ or more complex spin textures including skyrmions and skyrmioniums.

In this Letter we will provide answers to all these fundamental questions. We will show that the pattern of $\mathbf{D}_{ij}$  in low-dimensional itinerant magnetic structures grown on heavy-element metallic substrates can be rather complex. We introduce the magnon spectroscopy as a versatile tool to resolve such complex patterns of DMI. We provide direct experimental evidence for a change of the chirality index of DMI, which manifests itself in the asymmetry of the magnon dispersion relation. We will shed light on the origin of the observed effect and provide guidelines for quantum engineering of DMI in low-dimensional magnets on the atomic scale.

\begin{figure}[t!]
	\centering
	\includegraphics[width=.85\columnwidth]{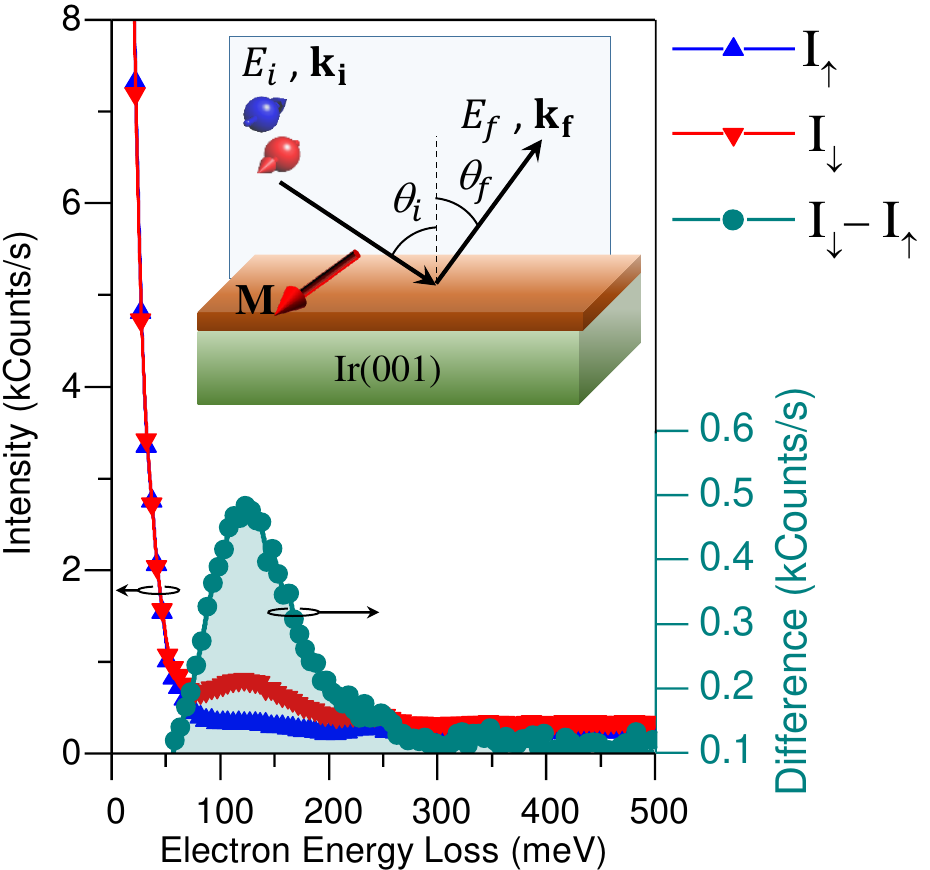}
	\caption{Typical SPHREELS spectra recorded at a wavevector of $Q=0.65$ \AA$^{-1}$ on a Co double layer epitaxially grown on Ir(001). The spectra are recorded at the incident energy of $E_i=8$ eV and at room temperature. The red and blue spectra, denoted by $I_{\downarrow}$  and $I_{\uparrow}$, are recorded with the spin polarization vector of the incident electron beam being parallel and antiparallel to the magnetization $\mathbf{M}$, respectively. The difference spectrum $I_{\downarrow}-I_{\uparrow}$ is shown by the sea-green color. The scattering geometry is schematically illustrated in the inset. The energy and wavevector of the incident (scattered) beam are shown by $E_i$ and $\mathbf{k_i}$  ($E_f$ and $\mathbf{k_f}$), respectively.}
	\label{Fig1:Spectra}
\end{figure}

\begin{figure}[t!]
	\centering
	\includegraphics[width=0.99\columnwidth]{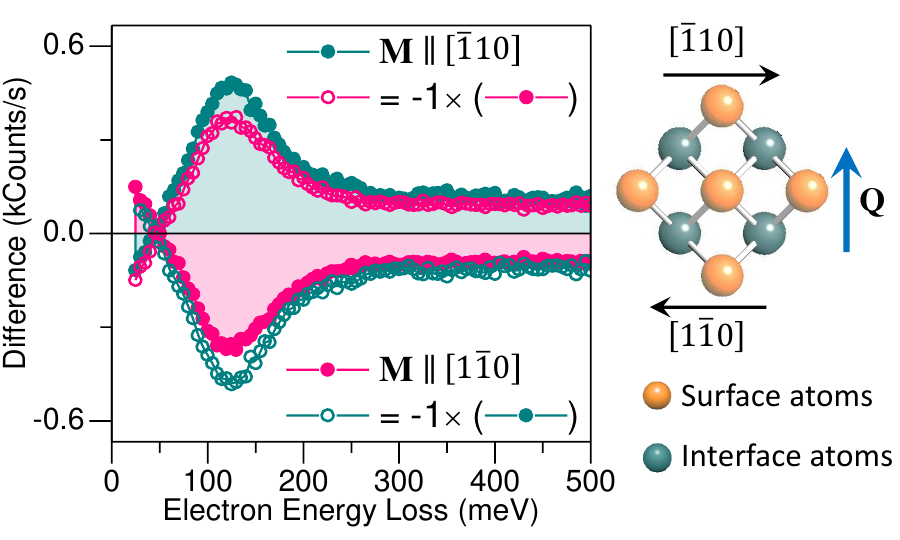}
	\caption{Difference spectra recorded at $Q=0.65$ \AA$^{-1}$ for the two opposite directions of magnetization, i.e.,  $\mathbf{M}\parallel[\overline{1}10]$ (sea-green solid circles) and $\mathbf{M}\parallel[1\overline{1}0]$  (red solid circles). In order to easily compare the spectra with different magnetization directions, the same spectra multiplied by $-1$ are shown as well. The magnon propagation direction (wavevector $\mathbf{Q}$) with respect to the principle directions of the Co layers and  $\mathbf{M}$ is schematically sketched in the right side.}
	\label{Fig2:Difference}
\end{figure}

We examine an epitaxial Co double layer on Ir(001) as a representative of 2D systems. All the experimental details are provided in Supplemental Material \cite{Note_Supp}.
We have shown earlier that the presence of DMI would lead to an asymmetry in the magnon dispersion relation \cite{Udvardi2009,Zakeri2010, Zakeri2012} and hence magnon spectroscopy provides a way to identify DMI \cite{Zakeri2017,Tsurkan2020}. The magnon dispersion relation was probed along the $\bar{\Gamma}$--$\bar{\rm X}$ direction of the surface Brillouin zone by means of spin-polarized high-resolution electron energy-loss spectroscopy (SPHREELS) \cite{Zakeri2012b, Zakeri2013, Zakeri2014b}. Typical SPHREEL spectra are presented in Fig.~\ref{Fig1:Spectra}. The spectra were recorded for different spin polarizations of the incoming electron beam. $I_{\downarrow}$ ($I_{\uparrow}$) represents the intensity spectrum when the spin polarization of the incoming electron beam was parallel (antiparallel) to the ground state magnetization $\mathbf{M}$. The difference spectrum $I_{\downarrow}-I_{\uparrow}$ provides all the necessary information regarding the magnons, e.g., their energy and lifetime \cite{Zhang2011, Zhang2012}. In this experiment $\mathbf{M}$ was parallel to the [$\overline{1}$10]-direction and the magnon wavevector $\mathbf{Q}$ was along the [$\overline{1}$$\overline{1}$0]-direction (see Figs.~\ref{Fig1:Spectra} and \ref{Fig2:Difference}).

DMI lifts the degeneracy of magnons having the same value of the wavevector but opposite propagation directions. Therefore, measuring the energy asymmetry $\Delta\varepsilon(\mathbf{Q})=\varepsilon(\mathbf{Q})-\varepsilon(-\mathbf{Q})$ of the magnon dispersion relation provides a way to quantify the strength of DMI \cite{Zakeri2010}. However, experimental determination of this asymmetry is not trivial, as probing the magnons with opposite orientations of $\mathbf{Q}$ requires a change in the scattering geometry, which may lead to unwanted effects. We have shown that a more accurate way is to perform a time-inversion experiment, keeping the scattering geometry unchanged \cite{Zakeri2012}. This can be realized by reversing the direction of  $\mathbf{M}$. The energy asymmetry can, therefore, be defined as $\Delta\varepsilon(\mathbf{Q})=\varepsilon_{\mathbf{M}\parallel[\overline{1}10]}(\mathbf{Q})- \varepsilon_{\mathbf{M}\parallel[1\overline{1}0]}(\mathbf{Q})$, where $\varepsilon_{\mathbf{M}\parallel[\overline{1}10]}(\mathbf{Q})$  and $\varepsilon_{\mathbf{M}\parallel[1\overline{1}0]}(\mathbf{Q})$ denote the magnon energy with the wavevector $\mathbf{Q}$ when $\mathbf{M}$ is parallel to the $[\overline{1}10]$- and $[1\overline{1}0]$-direction, respectively. Figure \ref{Fig2:Difference} shows the difference spectrum recorded for $Q=0.65$ \AA$^{-1}$ and for two different orientations of $\mathbf{M}$, i.e., $[\overline{1}10]$ and $[1\overline{1}0]$.

\begin{figure}[t!]
	\centering
	\includegraphics[width=0.99\columnwidth]{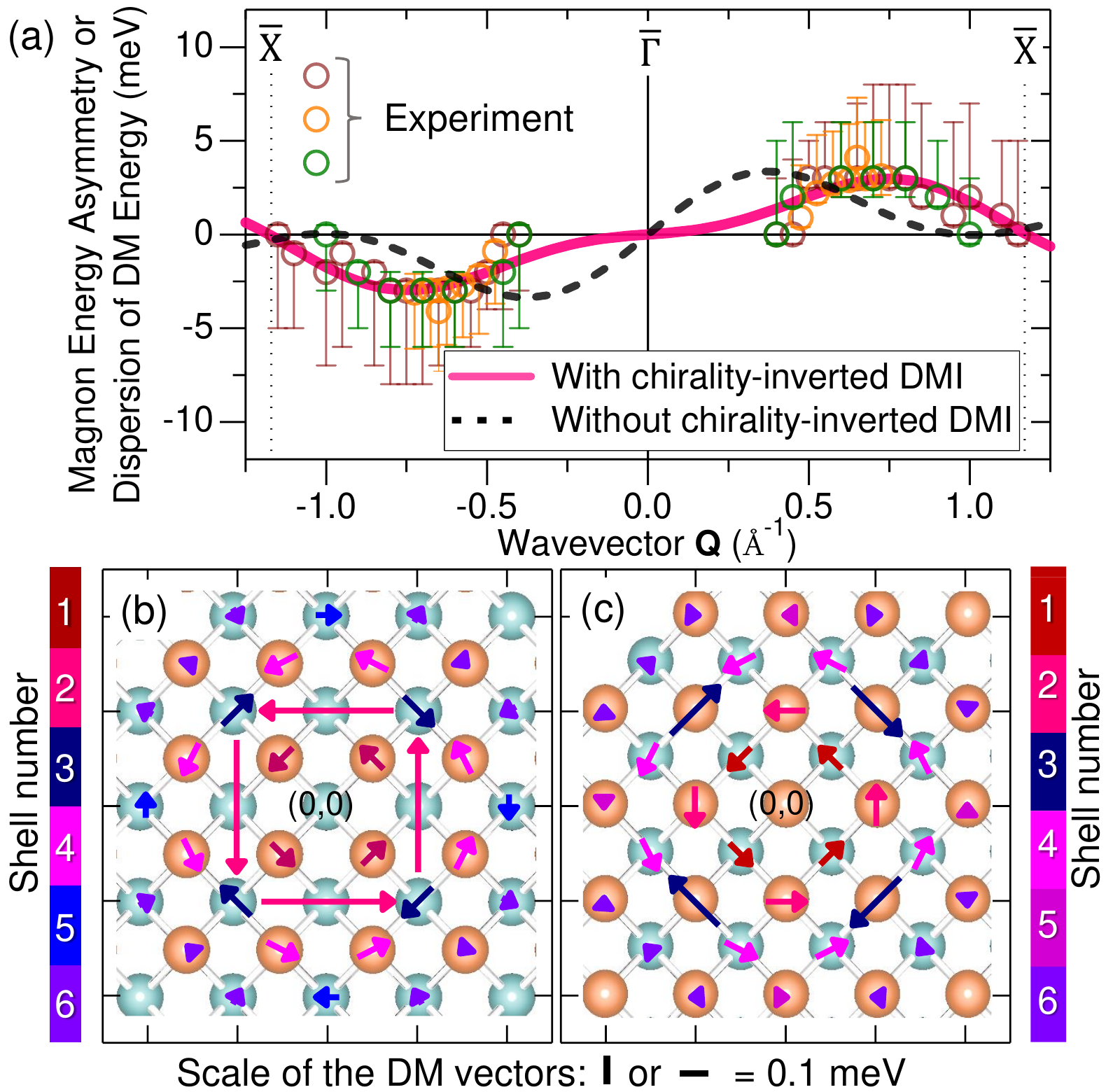}
	\caption{(a) Dispersion relation of the DM energy $\varepsilon(\mathbf{Q})$. The experimental data are shown by the open circles. Different colors indicate the results of different (but similar) samples. The error bars represent the standard deviations in the values of $\Delta\varepsilon$. The results of  \textit{ab initio} calculations are shown by the solid red curve. The black dashed curve indicates the artificial assumption of all DMI having alike counter clockwise chirality. (b) and (c) DM vectors from \textit{ab initio} calculations when the origin site (0,0) is located in the interface Co layer (b) and in the surface Co layer (b). In the ball representation of the Co atoms the sea-green and orange colors indicate the interface and surface layers, respectively. The color scales represent the order of nearest neighbors. The chirality index is also encoded in the color code. More red (blue) means a counter clockwise (clockwise) rotation is favored.}
	\label{Fig3:DMIDispersion}
\end{figure}

Data presented in Fig.~\ref{Fig2:Difference} unambiguously indicate that in this system DMI is present. However, the value of $\Delta\varepsilon$ is only about  $4$ meV, which seems to be rather small, at first glance. In order to shed light on the origin of this low value of $\Delta\varepsilon$ and also to further quantify the atomistic DM vectors we have probed the so-called dispersion relation of the DM energy, i.e., energy asymmetry versus wavevector $\Delta\varepsilon(\mathbf{Q})$.  A summary of the results from several different experiments performed on different samples (thicknesses of 1.8, 2 and 2.4 atomic layers)  is presented in Fig.~\ref{Fig3:DMIDispersion}. Since the quantity $\Delta\varepsilon(\mathbf{Q})$ is antisymmetric with respect to $\mathbf{Q}$, the data seem to be mirrored with respect to the origin. Note that at low wavevectors the dipolar scattering along with the large SOC of the system leads to interesting spin-dependent effects (a discussion on such effects is out of the scopes of the present Letter, see also Supplementary Note~II of \cite{Note_Supp}). We, therefore, focus on the spectra collected away from the dipolar scattering regime, where those effects are absent. Instead, we show the measurement with small steps of wavevectors.
%In the data set shown by the orange color the incident energy of the electrons used for magnon excitations was $E_i=8$ eV and in the other data set it was $E_i=9$ eV. Note that choice of $E_i$ does not influence the physics of the system. It merely changes the excitation cross-section and the signal to noise ratio.
Here the most important observations are: (i) the energy asymmetry is unexpectedly low, and (ii) more importantly, the maximum and the minimum of $\Delta\varepsilon(\mathbf{Q})$ are  located in the second half of the $\bar{\Gamma}$--$\bar{\rm X}$ path. Both observations are clear evidence of a chirality-inverted DMI (see below).

We resort to first-principles calculations of DMI, in order to gain further insight into the physics of these observations \cite{Note_Supp}. Our first-principles calculations are based on the fully relativistic Korringa-Kohn-Rostoker electronic structure method \cite{Ebert2011}. Details of the scheme used to compute the above quantities within the general framework of the magnetic force theorem \cite{Liechtenstein1987} are given in Ref.~\cite{Mankovsky2017}. The experimental interatomic distances were used as the input of the calculations  \cite{Heinz2009,Chen2017,Zakeri2021a,Qin2019,Zakeri2021}.

The calculations results are summarized in Figs.~\ref{Fig3:DMIDispersion}(b) and (c), where we show the DM vectors describing the antisymmetric interaction between the Co atoms only. The values are similar to the literature values for similar systems \cite{Freimuth2014, Hanke2018,Vida2016,Perini2018, Dupe2015, Heinze2011,Jadaun2020,Hoffmann2017,Zimmermann2019,Udvardi2009}. In Fig.~\ref{Fig3:DMIDispersion}(b) the interaction of a Co atom located at position $(0,0)$ within the interface layer, adjacent to the Ir(001) surface, with the other Co atoms at position $\mathbf{R}_{j}$ in the same layer (sea-green color)
or  located in the exposed surface layer (orange color) are shown. Similarly,  Fig.~\ref{Fig3:DMIDispersion}(c) displays the DMI between a Co atom located within the surface
layer at $(0,0)$ and the other Co atoms sitting at sites $\mathbf{R}_{j}$
either in the same layer (orange color)
or located in the deeper interface layer (sea-green color). The length of each vector represents the strength of the interaction $|\mathbf{D}_{i j}|$ and  their color represents their distance from the origin sites (0,0).  The rotation sense of the DM vectors is also encoded in their color. The red color means that a counter clockwise rotation is favored and the blue color indicates a clockwise rotation of the DM vectors. The chirality index is defined as $c=\mathbf{\hat{D}}_{ij}\cdot \mathbf{\hat{S}}_i \times \mathbf{\hat{S}}_j$, where the hat indicates the unit vector. Its sign is, therefore, directly given by the direction of $\mathbf{D}_{ij}$ \cite{Note1}.
Looking at the data presented in Fig.~\ref{Fig3:DMIDispersion}(b) one realizes that the first nearest neighbor DM vectors, shown in brown color are rather small ($D_{1,x}=D_{1,y}=-0.073$ meV). They exhibit a counter clockwise (CCW) rotation about the origin site (0,0) ($c<0$). The second nearest neighbor DM vectors, shown in red color, are rather large ($D_{2,x}=-0.45$, $D_{2,y}=0$) or ($D_{2,x}=0$, $D_{2,y}=-0.45$) meV and also show a CCW rotation sense ($c<0$). The surprising result was obtained for the third nearest neighbor (dark-blue color). The DM vectors show a clockwise (CW) rotation ($c>0$) with the components $D_{3,x}=D_{3,y}=+0.1$ meV. Interestingly, the  fourth nearest neighbors favor again a CCW rotation sense ($c<0$) but the fifth nearest neighbor DM vectors have the tendency to be of CW nature ($c>0$). Looking at the data presented in Fig.~\ref{Fig3:DMIDispersion}(c) one can draw a similar conclusion. The only difference is that the third nearest neighbor shows a more pronounced positive chirality (CW rotation)   $D_{3,x}=D_{3,y}=+0.16$ meV.  Unlike the interface layer the fifth nearest neighbor exhibits also a negative chirality index ($c<0$, CCW rotation).

Based on the values obtained from first principles, we calculated the dispersion relation of the DM energy and the results are shown by the solid red curve in Fig.~\ref{Fig3:DMIDispersion}(a) \cite{Note_Supp}. In line with the experiment one observes that the dispersion relation of the DM energy exhibits its extrema at $\pm 0.7$ \AA$^{-1}$. It is apparent that such an observation cannot be understood if all DMI terms possess the same chirality. This is demonstrated by the black dashed curve shown in Fig.~\ref{Fig3:DMIDispersion}(a). In such a case one must observe a rapid increase of $\Delta\varepsilon(\mathbf{Q})$ as $Q$ increases with  extrema located in the first half of the $\bar{\Gamma}$--$\bar{\rm X}$  symmetry direction, i.e., below $0.5$~\AA$^{-1}$ and a rapid decrease at larger wavevectors. In the region of small wavevectors $\Delta\varepsilon(\mathbf{Q})$
can be approximated by terms linear with respect to $Q$. Hence, only if the system exhibits a chirality inversion of DMI one would observe a reduction of the DM energy in this region, as a  result of competing terms with opposite signs. Note that the higher order Heisenberg type of exchange interactions are all of symmetric nature and hence do not appear in $\Delta\varepsilon(\mathbf{Q})$. Moreover, other interactions of chiral character are expected to be much weaker than DMI and their contribution to $\Delta\varepsilon(\mathbf{Q})$ may be neglected (see Supplementary Note II of \cite{Note_Supp}).

Our results clarify the long-standing question regarding the small micromagnetic DMI of the Co/Ir interface (cf. Refs. \cite{KloodtTwesten2019,Ishikuro2019} and references therein), which seems to be due to cancellation effect of negative and positive terms. The overall (micromagnetic) DMI is still negative (CCW rotation) \cite{Note_Supp}.

In order to shed light on the origin of the observed chirality-inverted DMI, one may start with the model by Fert and Levy for a three-sites interaction \cite{Fert1980,Levy1981}. According to this model $\mathbf{D}_{ij}$ between $\mathbf{S}_i$ and $\mathbf{S}_j$ mediated by a nonmagnetic atom sitting on site $n$ is given by
$\mathbf{D}_{ij}=\frac{D^0_{ij}}{R_{ij}} \sum_n\mathbf{R}_{in}\cdot\mathbf{R}_{jn}\left(\mathbf{R}_{in} \times \mathbf{R}_{jn}\right)/\left(R_{in}R_{jn}\right)^3$ \cite{Fert1980,Levy1981,Crepieux1998}. Here $\mathbf{R}_{in}$ and $\mathbf{R}_{jn}$ are the displacement vectors  and $D^0_{ij}=C\sin[\gamma(E_{\rm{F}})]\sin[k_{\rm{F}}(R_{in}+R_{jn}+R_{ij})+\gamma(E_{\rm{F}})]$, where $C$ is a constant proportional to the SOC strength and the strength of the interaction between the magnetic sites $i$ and $j$,  and $\gamma$ is the scattering phase shift of the electrons on site $n$, which, in turn, is related to the number of available $d$ electrons and the Fermi energy $E_F$ or  wavevector $k_F$. The vector product determines the direction of $\mathbf{D}_{ij}$. Considering the lattice, as shown in Figs.~\ref{Fig3:DMIDispersion}(b) and (c), the symmetry of the pairs of the magnetic atoms remains unchanged for all the neighbors. One would, therefore, expect the same sign for all $\mathbf{D}_{ij}$, as given by the vector product. For $D^0_{ij}>0$ one, would therefore expect a CCW rotation of $\mathbf{D}_{ij}$ and hence a negative chirality index ($c<0$). However, depending on the quantities appearing in the argument of sinus functions as well as on the sign of $C$ the prefactor $D^0_{ij}$ can either be positive or negative. Any change in the sign of $\mathbf{D}_{ij}$ must, therefore, be due to a sign change of $D^0_{ij}$. Conditions under which $D^0_{ij}$ can be positive or negative depend on the details of the electronic structures. Hence, it is not straightforward to provide simple guidelines for which system such an effect is expected (see also Supplementary Note III of \cite{Note_Supp}). Moreover, one has to recall that the coupling between Co atoms at sites $i$ and $j$ is mediated by electrons of surrounding atoms, both Co as well as Ir atoms, hybridized with the electrons of the interacting Co atoms \cite{Sandratskii2017}. The SOC-induced spin mixing as well as the itinerant nature of the electrons in the system are essential for the observed change in the chirality index of DMI via a mechanism similar to (but beyond) that of the RKKY interaction \cite{Ruderman1954,Kasuya1956,Yosida1957}. Interestingly, a similar  behavior has also been predicted for DMI in Co/Ir/Pt(111) \cite{Vida2016}, Ru/Co/Ir(111) and Mn/Re(0001) \cite{Meyer2020} and is expected to occur in many other combinations of $3d$ magnetic films, multilayers and nanostructures in contact with metallic substrates. Note that the observed chirality-inverted DMI reported here is different from the layer-dependent DMI discussed in Ref.~\cite{Yang2015} (see Supplemental Note II of \cite{Note_Supp}). Our analysis of $J_{ij}$ shows that there is no direct relation between the sign of $\mathbf{D}_{ij}$ and that of $J_{ij}$  \cite{Note_Supp}, in line with the results of Ref.~\cite{Vida2016}. Therefore, not only the overlap of the electronic wavefunctions but also the spin mixing and orbital contribution, as a consequence of SOC must be  important \cite{Kim2018,Jadaun2020}. All these effects are tightly connected to the degree of the hybridization of the electronic states of the magnetic atoms with those of the substrate atoms \cite{Kashid2014,Belabbes2016,Dupe2015}.

In conclusion, by probing the dispersion relation of the DM energy, we quantified the atomistic DMI in a model system of ultrathin ferromagnet on heavy-element substrate, i.e., a Co double layer epitaxially grown on Ir(001). Our detailed analysis of the DM energy dispersion showed that the pattern of the atomistic DMI in epitaxial magnetic structures can be very complex. Upon the increase of the interatomic distances DMI can change its  chirality index from positive to negative and vice versa, even though the symmetry of the system is unchanged. The effect is in analogy to the oscillatory HEI in ferromagnetic metals. The phenomenon was explained by comparing the experimental results to those of \textit{ab initio} density functional theory calculations and was attributed to the strong electronic hybridizations, the role of orbital degree of freedom and the presence of the spin-mixed itinerant electrons. The observed complex pattern of DM vectors is a general feature across different systems and is expected
to be present in many magnetic structures grown
on heavy-element metallic substrates (see Supplementary Note III \cite{Note_Supp}). Beside providing new insights into the microscopic origin of DMI, our results offer new routes to tune this fundamental interaction on the atomic scale. Moreover, our work showcases how magnon spectroscopy, which directly probes the dispersion of DM energy, can be used to quantify the atomistic DMI in great detail and experimentally identify any chirality-inverted DMI.

\section*{Acknowledgments}

Financial support by the Deutsche Forschungsgemeinschaft (DFG) through the DFG Grants ZA~902/7-1 and ZA~902/8-1 and the Heisenberg Programme (Grants ZA~902/3-1 and ZA~902/6-1) is acknowledged. Kh.Z. thanks the Physikalisches Institut for hosting the group and providing the necessary infrastructure. A.M. gratefully acknowledges partial financial support by the Czech Science Foundation grant GA~\v{C}R 23-04746S, by the Deutscher Akademischer Austauschdienst program ``Bilateral exchange of academics'', and computational resources by the IT4Innovation grant OPEN-24-35 ``CHIRSPIN''.

\bibliography {./Refs}

\end{document}